# Study of the spin-pump-induced inverse spin-Hall effect in Bi doped n-type Si

A.A.Ezhevskii[1]*, D.V.Guseinov[1], A.V.Soukhorukov[1], A.V.Novikov[2], D.V.Yurasov[2], N.S.Gusev[2]

[1]Lobachevsky State University of Nizhny Novgorod, 23 Pr. Gagarina (Gagarin Avenue), 603950 Nizhny Novgorod, Russia.

[2]Institute for Physics of Microstructures Russian Academy of Sciences, Academicheskaya Str., 7, Afonino, Nizhny Novgorod region, Kstovsky district, 603087, Russia.

*ezhevski@phys.unn.ru

An inverse spin Hall effect (ISHE) in n-type silicon was observed experimentally when conduction electrons were scattered on the spin-orbit potential of bismuth. The spin current in the silicon layer was generated by excitation of the magnetization precession during ferromagnetic resonance in a thin permalloy (Py) layer deposited on a Si layer doped by phosphor and bismuth. From the angular dependences of the dc voltage for different Py/n-Si:Bi structures aligned along the [011] or [100] crystal axes, we were able to distinguish the planar Hall effect (PHE) and ISHE contributions. The ISHE dc voltage signal was proportional to $sin\theta \cdot sin 2\theta$ product for the structure aligned to the [011] crystal axis and to $sin\theta \cdot cos 2\theta$ for the [100] direction. In addition, the PHE dc voltage was observed for the angles corresponded to the $sin 2\theta$ dependence. It means that for silicon as a many-valley semiconductor, the scattering of spins due to the spin-orbit potential induced by shallow donor in n-type material is dependent on the orientation of the valley axes relative to the direction of the magnetic field.

## INTRODUCTION

There has been a growing theoretical and experimental interest focused on the spin-Hall effect (SHE), which refers to the generation of spin currents from charge currents via the spin-orbit interaction [1–7]. The spin-orbit interaction responsible for the SHE may also cause the inverse spin-Hall effect (ISHE), which is the process that converts a spin current into an electric voltage [8–11]. The ISHE was observed and investigated in simple metallic ferromagnetic/paramagnetic $Ni_{81}Fe_{19}$/Pt bilayer systems using a spin-pumping method operated by ferromagnetic resonance (FMR) [8-11]. The spin pumping driven by ferromagnetic resonance injects a spin current into the paramagnetic layer, which gives rise to an electromotive force transverse to the spin current using the ISHE in the paramagnetic layer. In a $Ni_{81}Fe_{19}$/Pt film, an electromotive force was found perpendicular to the applied magnetic field at the ferromagnetic resonance condition. The electromotive force was observed also in a Pt/$Y_3Fe_4GaO_{12}$ film, in which the metallic ferromagnetic layer was replaced by an insulating $Y_3Fe_4GaO_{12}$ layer, supporting that the spin-pumping-induced ISHE is responsible for the observed electromotive force [11]. In [12] the ISHE was studied in p-type silicon by holes scattering on the lattice spin-orbit potential (SOP).

A specific feature of n-type silicon which is considered as a material for spintronic future applications is a weak spin-orbit interaction for electrons, which on the one hand leads to a weak spin accumulation effects due to the small angles of the spin Hall effect, and on the other hand - to significantly lower spin relaxation rates and long spin diffusion length. Strong spin-dependent scattering on the spin-orbit potential (SOP) can be induced by doping of Si by heavy donor of group V. The latter could make it possible to control the spin-orbit contribution to the scattering, which may lead to the generation and detection of spin currents under certain conditions.

In this paper, we report the observation of ISHE in n-type silicon for the first time. In this case conduction electrons which were spin-polarized by ferromagnetic resonance in the permalloy layer deposited on silicon are scattered on the spin-orbit potential of bismuth. Due to the small value of ISHE in n-type silicon, some additional effects such as anomalous Hall effect (AHE) and anisotropic magnetoresistance (AMR) have to be considered and compared with the ISHE.

**EXPERIMENTAL**

To study the effects associated with the excitation and detection of spin current in n-type silicon the phenomenon of spin pumping [1] was exploited. The investigated structure was formed by the following procedure. At first nominally undoped Si layer was deposited on the commercial SOI wafer (provided by Soitec) by molecular beam epitaxy. The residual doping level was of the order of $10^{15}$ cm$^{-3}$ (p-type). After that this layer was doped by P and Bi via ion implantation (Bi: $\Phi$=0,005 μC/cm$^2$, $E$=24 keV; P: $\Phi$=600 μC/cm$^2$, $E$=100 keV) and subsequent annealing (1000°C, 30 min). The Au/Ti (1.0×0.5mm$^2$) ohmic contacts and a Py layer (1.0×1.8mm$^2$) were formed on Si layer (1.0×3.0 mm$^2$) using the lift-off lithography and magnetron sputtering. The Py/n-Si:Bi structures with two different orientations (the longer side of structure was aligned to the [101] or [100] crystal axis) were formed. Usage of SOI substrates allowed eliminating the impact of substrate conductivity on the obtained results. The structure layout is shown in Figure 1a and a schematic illustration of spin pumping is shown in Figure 1b.

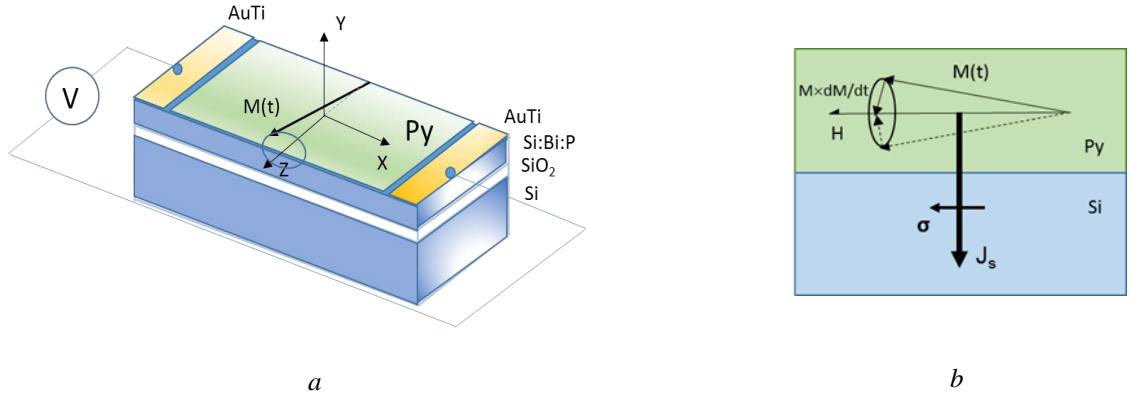

*a*                          *b*

Figure 1.*a* - Schematic view of the Py/n-Si:Bi structure for the ISHE study. Precession of the magnetization direction **M**(*t*) of the ferromagnetic Py layer pumps spins into the adjacent Si:Bi:P layer by inducing a spin current; *b* - schematic illustration for the pure-spin-current injection **J**$_s$ by spin pumping (H – magnetic field, $\sigma$ - the spin-polarization vector).

In order to proof the bismuth involvement in the ISHE spin scattering, the Py/Si:P structure without Bi doping was made as a reference sample. Phosphorus doping of silicon layer was carried out at the same conditions as for the structures with Bi doping.

FMR spectra in Py layer were measured at room temperature with the Bruker_EMX-plus-10/12 electron spin resonance spectrometer operating at 9.4 GHz microwave (MW) frequency. The Py/Si:Bi/SOI structure was placed near the center of a high-Q TE$_{011}$ cylindrical microwave cavity at which the magnetic component of the microwave field (*h*) is maximized while the electric-field component is minimized. The voltage on the Au/Ti contacts was measured using a lock-in amplifier at a modulation frequency of the magnetic field of 100 kHz and was recorded using an oscillograph.

**PRECESSION-INDUCED SPIN PUMPING IN Py/n-Si:Bi FILM**

To describe the injection of spins from a thin layer of ferromagnetic material into the n-type silicon using the spin pumping method, one can use the theoretical model proposed in [14, 15] for a ferromagnetic–normal metal (F-N) structure. As it is well known an F-N interface leads to a dynamical coupling between the ferromagnetic magnetization and the spins of the conduction-band electrons in the normal metal [16, 17]. When the magnetization direction precesses under the influence of an applied magnetic field, a spin current *I*$_s$ is pumped out of the ferromagnetic [15] can be written as:

$$I_s^{pump} = \frac{\hbar}{4\pi} g_r^{\uparrow\downarrow} \left[ \vec{m}(t) \times \frac{d\vec{m}(t)}{dt} \right]. \qquad (1)$$

Here, **m**(*t*) is a unit vector of the time-dependent magnetization direction of the ferromagnetic which at time *t* rotates around the vector of the magnetic field **H**$_{\text{eff}}$ (**m**(*t*)⊥**H**$_{\text{eff}}$) and $g_r^{\uparrow\downarrow}$ is the real part of the mixing conductance, which will be determined bellow. This follows the conservation of energy $\Delta E_F = -\Delta E_N$ and angular momentum $\Delta L_F = -\Delta L_N$, according to which the number of spins pumped from ferromagnetic into the normal metal (being equal to $N_s$) transfers the energy $\Delta E_N = N_s \mu_s/2$ and the angular momentum $\Delta L_N = N_s \hbar/2$. Here $\mu_s$- the spin accumulation or nonequilibrium chemical potential imbalance, $\Delta E_F = g \Delta L_F H_{\text{eff}}$ - the magnetic energy, and g is the gyromagnetic ratio of the ferromagnetic. The spin current from (1) leads to a damping of the ferromagnetic precession, resulting in an alignment of the magnetization with the applied magnetic field.

From the microscopic point of view the spins **s** of the conduction electrons in the normal conductor are coupled to the localized spins **S** of 3d magnetic electrons in the ferromagnetic by the s-d exchanged interaction in the vicinity of the F-N interface, and this can be described by the term $2J_{sd}\mathbf{s}\cdot\mathbf{S}$.

Following the concept described above in FMR-induced spin pumping experiments used for Py/n-Si:Bi system, an external magnetic field **H** and a microwave field **h** are applied and a precession motion of the magnetization **M** is excited by absorbing the angular momentum from the microwave at the resonance condition (see Figure 1). The magnetization precession damping proportional to **M**×d**M**/*dt* produces a pure spin current. The time average of **M**×d**M**/*dt* generates a dc pure spin current **J**$_s$ carrying the spin-polarization vector $\sigma$.

Here, the steady magnetization precession is maintained by balancing the absorption and the emission of the angular momentum of the magnetization. This emission is proportional to $\mathbf{M}(t)\times\frac{d\mathbf{M}(t)}{dt}$ term in the Landau–Lifshitz–Gilbert equation [16]:

$$\frac{d\mathbf{M}(t)}{dt} = -\gamma\mathbf{M}(t)\times\mathbf{H}_{\text{eff}} + \frac{\alpha}{M_s}\mathbf{M}(t)\times\frac{d\mathbf{M}(t)}{dt}, \qquad (2)$$

Here, γ, α, and M$_s$ are the gyromagnetic ratio, the Gilbert damping constant, and the saturation magnetization, respectively.

When our semiconductor layer is connected to the ferromagnetic (see Fig. 1b), the spin-polarization propagates into the semiconductor, which gives rise to a pure spin current with a spatial direction *j*$_s$ along *y* axis and a spin-polarization direction σ||**H**.

According to [1, 4, 5] (see expressions (1) and (2)) the precession of the magnetization direction **M** is caused by the torque ∝ **M**×**H**$_{\text{eff}}$ and the direct-current component of a generated spin current at the Py/n-Si:Bi interface (*y*=0) is expressed as:

$$\mathbf{j}_s^0 = \frac{\omega}{2\pi}\int_0^{2\pi/\omega} \frac{\hbar}{4\pi} g_r^{\uparrow\downarrow} \frac{1}{M_s^2}\left[\mathbf{M}(t)\times\frac{d\mathbf{M}(t)}{dt}\right]_z dt = \frac{g_r^{\uparrow\downarrow}\gamma^2 h^2 \hbar\left[4\pi\pi_s\gamma + \sqrt{(4\pi 4_s)^2\gamma^2 + 4\omega^2}\right]}{8\pi\pi^2\left[(4\pi 4_s)^2\gamma^2 + 4\omega^2\right]}, \quad (3)$$

where $\hbar$ is the Dirac constant and the real part of the mixing conductance $g_r^{\uparrow\downarrow}$ is given in [5] as:

$$g_r^{\uparrow\downarrow} = \frac{2\sqrt{3}\pi M_s \gamma d_F}{g\mu_B \omega}(\Delta H_{pp}^{F/N} - \Delta H_{pp}^F). \quad (4)$$

Here $d_F$ – the thickness of the Py film; $\omega$ and $\gamma$ are the angular frequency of magnetization precession and the gyromagnetic ratio, respectively; $\Delta H_{pp}^{F/N}$ and $\Delta H_{pp}^F$ are the FMR spectral width for the Py/Si:Bi:P layer and the Py film, respectively which are proportional to the Gilbert damping constant α:

$$\Delta H_{pp} = (2\omega/\sqrt{3}\gamma)\alpha, \quad (5)$$

In the Py/n-Si:Bi structure the spin current injected into the Silayer decays along the *y* direction (see Fig. 1) due to spin relaxation. According to [13] it can be expressed as:

$$j_{s(y)} = \frac{\sinh[(d_N - y)/\lambda_N]}{\sinh(d_N/\lambda_N)} j_s^0 \quad (6)$$

Here $d_N$ is the thickness and $\lambda_N$($d_N>\lambda_N$) is spin-diffusion length in the uniformly doped n-Si:Bi:P layer.

**SPIN CURRENT CONVERSION TO A CHARGE CURRENT BY ISHE INPy/n-Si:Bi FILM**

Scattering of polarized spins of the spin current on the SOP of Bi atoms in n-Si layercause the inverse spin-Hall effect that converts a spin current into a charge current [8]:

$$\vec{j}_C = \frac{e\theta_{SH}}{\hbar}\left[\vec{J}_S \times \vec{\sigma}\right] \quad (7)$$

Here $\vec{\sigma}$ denotes the spin polarization vector of the spin current $\vec{J}_S$ (see Figure 1), $\theta_{SH}$ is the spin-Hall angle.

Using Eq. (6) and (7), one can obtain the averaged charge current density

$$\langle j_c \rangle = \frac{1}{d_N} \int_0^{d_N} j_c(y)\,dy \tag{8}$$

$$\langle j_c \rangle = \theta_{SHE}\left(\frac{2e}{\hbar}\right)\frac{\lambda_N}{d_N} tanh\left(\frac{\lambda_N}{2d_N}\right) j_S^0 \tag{9}$$

and for dc ISHE voltage one can write [11]:

$$V_{ISHE} = \frac{w e \theta_{SH} \lambda_N\, tanh(d_N/2\lambda_N)}{d_N \sigma_N + d_F \sigma_F}\left(\frac{2e}{\hbar}\right) j_S^0 \tag{10}$$

Here $w$ is the width of the Py layer, $d_N$ and $\lambda_N$ – the thickness and the spin-diffusion length of the nonmagnetic layer and the $\sigma_F$ and $\sigma_N$ are the electrical conductivity of the Py and n-Si:Bi layers respectively.

In all our experiments, the static magnetic field lied in the plane of the structure and microwave magnetic field was orthogonal to this plane as shown in Figure 2.

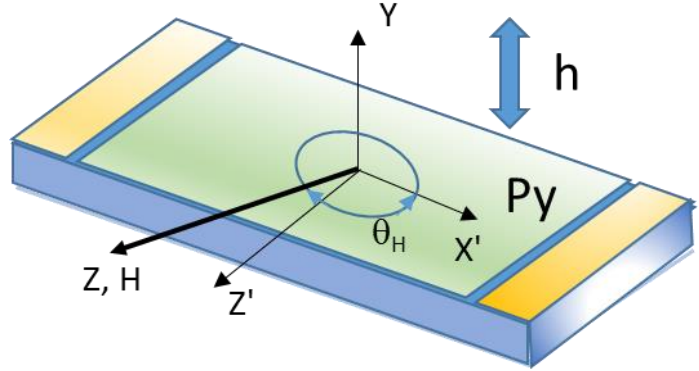

Figure 2. Configurations of the magnetic fields related to the plane of the Py/n-Si:Bi layer: the static magnetic field $H$ rotates in the plane of the structure and microwave magnetic field $h$ is orthogonal to this plane.

Here the x' and z' axes of the Cartesian coordinate system (x′, y, z′) are directed along the sample length and width while the y axis is the normal to the sample surface. The coordinate system (x, y, z) is chosen so that the z axis is parallel to the applied magnetic field H.

Detection of the ISHE was carried out by measuring the dc voltage signal appearing between the Au/Ti contacts (see figure 1). For such configurations of magnetic fields and magnetization, one may not take into account the contribution of the anomalous Hall effect [18] but in addition to ISHE it is necessary to consider the transverse part of the anisotropic magnetoresistance (AMR) in permalloy, which is known as planar Hall effect (PHE) [18] for the in-plane configuration of the static magnetic field. In this case, the microwave current, which is coupled to the microwave resistance, can produce a dc electromotive force and voltage:

$$\vec{E}_{PHE} = \frac{\Delta\rho}{M^2}(\vec{J}\cdot\vec{M})\vec{M} \qquad (12)$$

$$V_{PHE}(\theta) = \langle Re\{I(t)\}\cdot Re\{H(t)\}\rangle_{2\pi/\omega} = \\ = \frac{1}{2}I_0 h_0 \nabla R(H_0)\cos\Phi = \Delta R_{PHE} I_0 h_0 \sin 2\theta \cos\Phi \qquad (13)$$

Here $\Phi$ is a phase shift associated with the losses in the system and the resistance is expanded in a Taylor series around the dc field $H_0$ to first order in the microwave field $h(t)$, $R(H(t)) = R(H_0) + h(t)\nabla R(H_0)$[18].

The angular dependences of AMR are summarized in [18] for each measurement configuration. For our case one can see that $V_{PHE}\propto\sin 2\theta$. As suggested in [18] the only way to distinguish between ISHE and PHE signals is to study the angular dependences of the dc voltage signal on the direction of the magnetic field.

Silicon is a many-valley semiconductor. After scattering, the conduction electron transferred to a valley on a different crystal axis in $k$-space. Since the valleys have a magnetic anisotropy ($g_\parallel \neq g_\perp$), scattering of spins due to the spin-orbit potential induced by shallow donor in n-type material may be dependent on the orientation of the valley axes relative to the direction of the magnetic field. If we consider the transitions between different valleys, taking into account the spin conservation ($M_S$=const) in SHE scattering, then one could see (Figure 3) that not all transitions are allowed due to the anisotropy of the valleys ($g_\parallel \neq g_\perp$).

This can be understood by considering the SO interaction which can be written as:

$$\hat{H}_{SO} = \lambda(\hat{L}_Z\hat{S}_Z + \hat{L}_X\hat{S}_X + \hat{L}_Y\hat{S}_Y) = \lambda\left[\hat{L}_Z\hat{S}_Z + \frac{1}{2}(\hat{L}_+\hat{S}_- + \hat{L}_-\hat{S}_+)\right] \qquad (14)$$

We will consider the matrix elements of transitions between different band (valley) states, which consists of six degenerate basis states ($A_1$, $E$ and $T_2$). The only sufficient non-vanishing matrix elements of the orbital momentum between these states are:

$$\langle T_Y|\hat{L}_X|T_Z\rangle = \langle T_Z|\hat{L}_Y|T_X\rangle = \langle T_X|\hat{L}_Z|T_Y\rangle = -i \qquad (15)$$

Accordingly, for SHE processes the effective spin-orbit interaction involves only the $\langle T_X|\hat{L}_Z|T_Y\rangle$ term, because for another terms the spin operators $\hat{S}_X, \hat{S}_Y$ are responsible for spin-flip processes and not allowed for SHE scattering. As it is shown in Figure 3 the main axes of the Z'' and X'' valleys are directed at an angle of 45° relative to Z'||[10-1] axis in the (010) plane of the layer. In this case, the transitions involved in the SHE occur only when the magnetization is orthogonal to the main axes of the Y and X'' valleys (H||Z'') and the ISHE signal is zero when H||Z'.

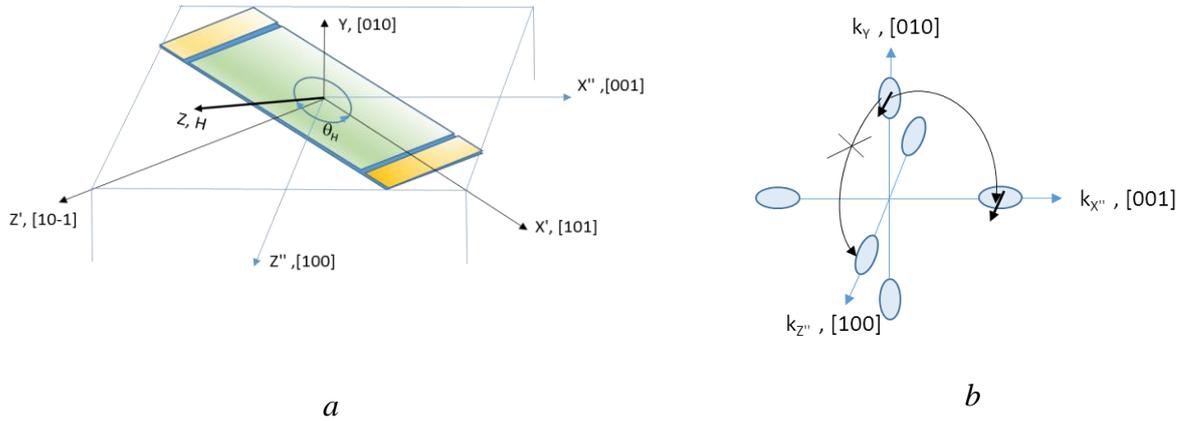

*a*  *b*

Figure 3. a) - the Py/n-Si:Bi structure with the structure aligned to the [101] crystal axis. *b*- allowed (Y→X'' valley) and not allowed (Y→Z'' valley) transitions for spin scattering at the SHE(see text).

Therefore, the transitions between different valley states give the additional $sin2\theta$ angular dependence to the general dependence of $sin\theta$ [11], and so, for the ISHE dc voltage one can write that $V_{ISHE} \propto sin\theta \cdot sin2\theta$ dependence.

**EXPERIMENTAL RESULTS AND DISCUSSIONS**

Earlier studies [19] of spin resonance spectra of conduction electrons in n-type Si showed that spin scattering of conduction electrons on the spin-orbit potential of Bi makes a significant contribution to their linewidths and g-factors. The suggestion to observe the ISHE experimentally in silicon doped with heavy donors as a small contribution to ordinary Hall effect was made in [20]. According to [19] bismuth in silicon can lead to dramatic increasing of spin-relaxation rate, which may suppress the ISHE. For that in spin-pumping experiment we were forced to use rather low bismuth concentration ($10^{16}$cm$^{-3}$ or lower). In order to enhance the spin pumping effect and spin current we have implemented the heavy (>$10^{19}$cm$^{-3}$) P doping.

Figure 4a shows the angular dependence of the dc voltage signal for the rotation of the magnetic field in the plane of the layer. As can be seen the signal has a maximum when the angle between magnetic field and Z' (see Figure 3) axis is θ=45° and no signal was found when the direction of H was parallel to Z' axis. For observation of ISHE it was unusual, because for Py/Pt, and Py/p-type Si layers (see ref. [1,2]) the ISHE dc voltage signal has a maximum at the direction of the magnetic field H||Z' ($\theta$=90° in Figure 3), and described by $sin\theta$ dependence for the in-plane rotation of the static magnetic field [6]. On the other hand, this dependence does not completely coincide with the angular dependence for the PHE voltage proportional to $sin2\theta$.

We suggested that the observed angular dependences of the dc voltage signal contain both PHE and ISHE contributions: $V=V_{ISHE}+V_{PHE}=asin\theta sin2\theta+bsin2\theta$. The angular dependences of the dc voltage signal, containing PHE and ISHE contributions for Py/n-Si:Bi structure in which the structure is aligned to the [101] crystal axis are shown in Figure 4a and b.

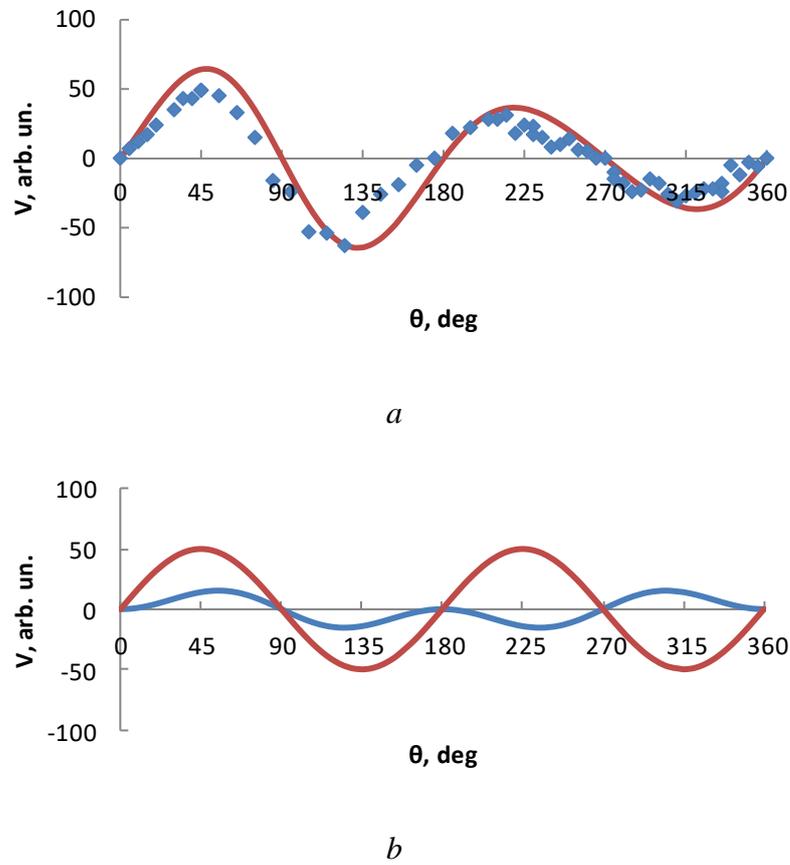

*a*

*b*

Figure 4. The angular dependences of the dc voltage, containing PHE and ISHE contributions for Py/n-Si:Bi structure in which the structure is aligned to [101] axis. *a*- experimental data (diamonds) and modeling (solid line) which takes into account both contributions ($V=V_{ISHE}+V_{PHE}$); *b* - theoretical approximation of the ISHE (blue line) and PHE (red line) ($V_{ISHE} = a\sin\theta \cdot \sin 2\theta$ and $V_{PHE} = b\sin 2\theta$, a/b=0.4).

To verify our hypothesis about the valley contribution to the SHE in n-type silicon, the structure aligned to [001] crystal axis was studied (Figure 5).

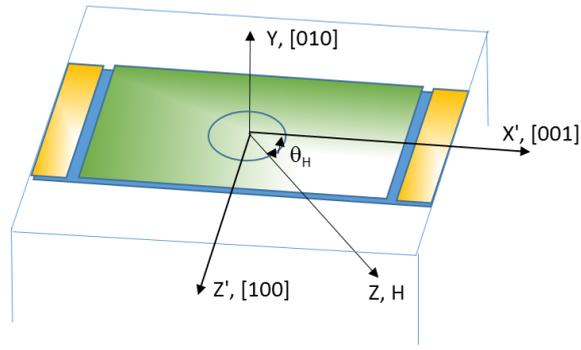

Figure 5. The Py/n-Si:Bi structure aligned to the [001] crystal axis.

The angular dependences of the dc voltage, containing PHE and ISHE contributions for such Py/n-Si:Bi structure are shown in Figure 6. In this case the ISHE dc voltage signal is proportional to $sin\theta \cdot cos2\theta$, which results in the fact that ISHE signal could be observed at θ=90 and 270° in addition to the PHE dc voltage at the angles corresponding to the $sin2\theta$ dependence. This followed to the valley involvement to the SHE scattering, when magnetization is orthogonal to the main axes of the Y and X' valleys (Figure 5). In this case the additional $cos2\theta$ multiplier is added to the $sin\theta$ dependence.

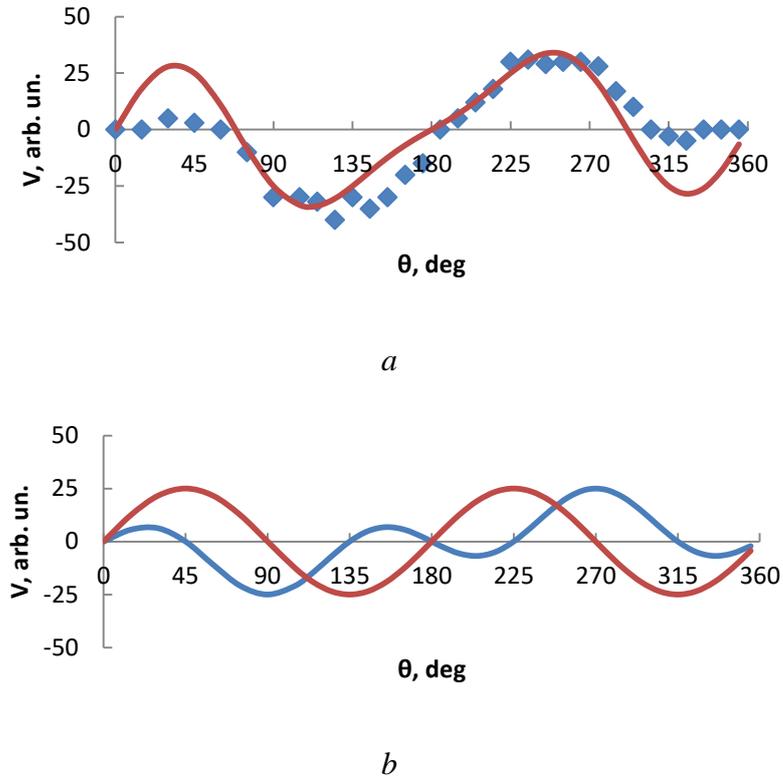

Figure 6. The angular dependences of the dc voltage, containing PHE and ISHE contributions for Py/n-Si:Bi structure in which the structure is aligned to the [001] crystal axis. *a*- experimental data (diamonds) and modeling (solid line) which takes into account both contributions

($V=V_{ISHE}+V_{PHE}$); b- theoretical approximation of the ISHE (blue line) and PHE (red line) ($V_{ISHE} = a' \sin\theta \cdot \cos 2\theta$ and $V_{PHE} = b' \sin 2\theta$, $a'/b'=1$).

Using the experimental parameters and the obtained results, we evaluated the angle $\theta_{SHE}$ of the spin Hall effect. Taking $\sigma_f = 1.49 \cdot 10^6 (\Omega \cdot m)^{-1}$, $\sigma_n = 5.0 \cdot 10^4 (\Omega \cdot m)^{-1}$, $V_{ISHE}=200$ nV (value corresponded to the maximum in the dependence for ISHE in figure 5b) for 200 mW microwave power, $4\pi M_s=0.99708$ T, $W_{Py/Si:Bi}-W_{Py} =0.6$ mT, $\alpha=0.00893$, $g^{\uparrow\downarrow}=4.450 \cdot 10^{17}$ m$^{-2}$, $j_s=6.2 \cdot 10^{-11}$ J/m$^2$, we have determined this angle to be $\theta_{SHE} \approx 0.0001$. This value coincides with the calculations made in [7] and is of the same order that was found in p-type silicon [12] and in gallium arsenide [7].

Since the spin diffusion length for n-type Si is much longer than for p-type silicon, it is still of great interest to obtain the Si:Bi:P layers with higher thickness to obtain a stronger ISHE dc voltage signal. Nevertheless, in the process of Bi doping of silicon, we faced the problem of obtaining the uniformly doped layers with a thickness higher than half of a micron, since we used the method of ion implantation for Bi and P doping, which limited the thickness of the doped layer.

We estimated the value of the PHE contribution to the dc voltage taking into account the known AMR ratio of $\Delta R/R(0)$ ~0.4% for permalloy layer [18]. Using parameters of our structure and microwave cavity, we could estimate that dc voltage on the edges of Py layer may be of order of 20 nV and slightly smaller value between the Au/Ti contacts (here we also not included some power loss in microwave cavity). This value is almost one order of magnitude smaller than we have measured in our experiment.

The difference can be understood on the basis of spin-orbit interaction contribution to the value of $\Delta R$ [21]. In the case of dynamic magnetization in the bilayer systems such as Py/n-Si:Bi both layers should be taken into account.

In order to make measurements of the PHE without the ISHE involvement to the dc voltage, the structure of Py/n-Si was made without bismuth doping. Phosphorus doping of the silicon layer was carried out by ion implantation and annealing at the same conditions as for structures with bismuth.
The angular dependences of the dc voltage, containing PHE contributions for Py/n-Si structure are shown in Figure 7.

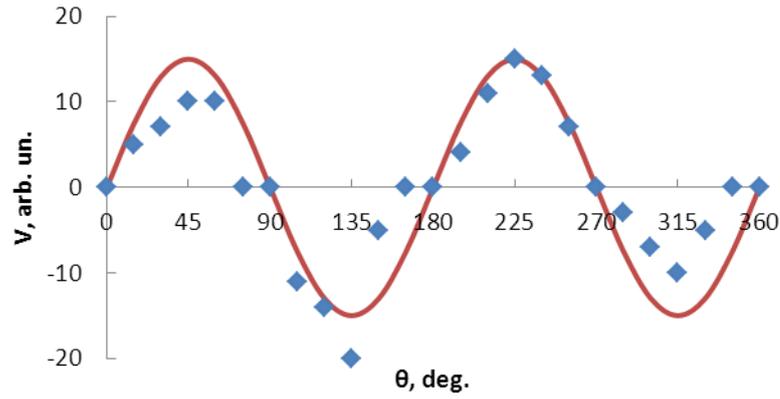

Figure 7. The angular dependences of the dc voltage, containing PHE contributions for Py/n-Si structure doped only with phosphorus ($N_P > 10^{19}$ cm$^{-3}$): experimental data (diamonds) and modeling (solid line) taking into account the PHE contribution ($V_{PHE} \propto \sin 2\theta$). The structure is aligned to the [001] crystal axis.

As it can be seen, only the PHE dc voltage signal is observed with angular dependence proportional to $\sin 2\theta$. Again, the magnitude of the PHE is much higher than was estimated from parameters of Py film, but smaller than that observed in the Bi doped structures. For Si doped with phosphorus the spin scattering due to the spin-orbit potential induced by phosphorus is much smaller than for silicon doped with bismuth. However, owing 3÷4 orders of magnitude higher concentration of phosphorus as compared to bismuth, it can lead to high spin relaxation rate which may contribute to the damping coefficients in LLG equation (2) for FMR. In this case one may suggest that contribution of spin-orbit interaction may come from n-Si layer and interface layer between Py and n-Si layers. It is consistent with FMR linewidths behavior for different structures like Py/n-Si:Bi, Py/n-Si:P and Py on high resistivity Si (Fig. 8).

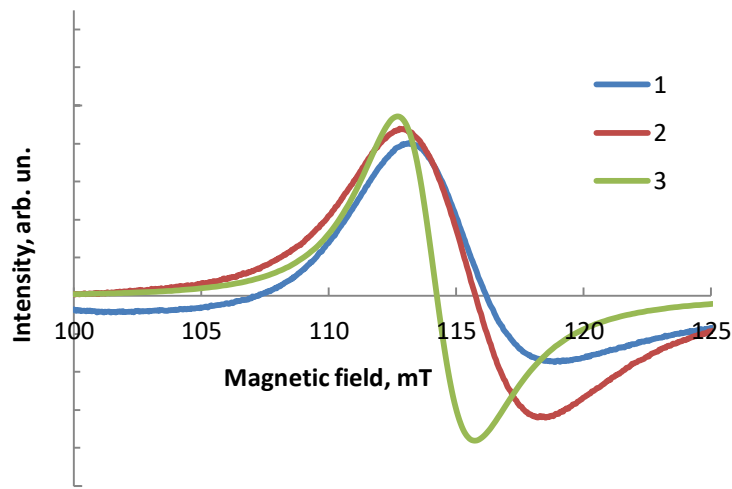

Figure 8. The FMR spectra of Py for different structures: 1 - Py/n-Si:B**i** ; 2 - Py/n-Si:P; 3 – Py on Si of high resistivity (5000 Ω cm).

It is seen that FMR linewidths for Py/n-Si:Bi and Py/n-Si:P are almost the same and they are remarkably higher than that for Py deposited on high resistivity Si. Such behavior of FMR line widths and PHE for Py/n-Si bilayer systems was not considered in literature earlier and requires some additional experimental studies and theoretical analysis.

## SUMMARY

In summary, we have studied the inverse spin Hall effect that was induced in n-type silicon due to scattering of the spin-polarized conduction electrons on the spin-orbit potential of bismuth using the spin-pumping effect. The spin current in the silicon layer was generated by excitation of the magnetization precession at ferromagnetic resonance in the thin Py layer deposited on n-Si layer doped by bismuth. From the angular dependences of the dc voltage for Py/n-Si:Bi structures with different orientation relative to crystal axes the planar Hall effect (PHE) and ISHE contributions was evaluated. It was obtained that the ISHE dc voltage signal was proportional to $sin\theta \cdot sin2\theta$ for the structure which was aligned to the [011] crystal axis and to $sin\theta \cdot cos2\theta$ for structure aligned the [100] one. In addition, the *sin2θ* dependence of the PHE dc voltage was observed. This leads to the contribution of the intervalley scattering to the SHE, and this gives an additional *sin2θ* or *cos2θ* factor to the usual *sinθ* dependence. It means that for silicon as a many-valley semiconductor, the scattering of spins due to the spin-orbit potential induced by shallow donor in n-type material is dependent on the orientation of the valley axes relative to the direction of the magnetic field.

Using the obtained results, the angle of the spin Hall effect was estimated from the magnitude of the ISHE voltage in the Py /n-Si:Bi structures to be $\theta_{SHE}$~0.0001. The value of the PHE contribution to the dc voltage is almost one order of magnitude smaller than we measured in our experiment. The difference can be understood on the basis of spin-orbit interaction contribution to the value of ΔR which in case of dynamic magnetization in the bilayer systems such as Py/n-Si:Bi can be determined by both layers.

Our results can be helpful for understanding some aspects of the spin-orbit interaction in semiconductors and for the engineering of Si-based spintronic devices.

## ACKNOWLEDGMENTS

We are grateful to Prof. A.A. Fraerman and Dr. E.A. Karashtin (Institute for Physics of Microstructures of the Russian Academy of Sciences) for interest to this work and stimulating discussions. Sample fabrication was supported by the RSF Grant№16-12-10340. Ferromagnetic resonance and spin pumping experiments were supported by RFBR Grant№18-03-00235-a.